\documentclass[trackchanges]{aastex701}

\begin{document}

\title{Nancy Grace Roman Space Telescope Wide Field Instrument: Grism Saturation Response to Bright Sources during Thermal-Vacuum Testing}

\author[orcid=0009-0000-9825-9755]{Lipika Chatur}
\affiliation{Steward Observatory, University of Arizona, 933 North Cherry Avenue, Tucson, AZ 85721, USA}
\email[show]{lchatur@arizona.edu}  

\author[0000-0002-2457-272X]{Dana R. Louie}
\affiliation{Catholic University of America, Department of Physics, Washington, DC, 20064, USA}
\affiliation{Exoplanets and Stellar Astrophysics Laboratory (Code 667), NASA Goddard Space Flight Center, Greenbelt, MD 20771, USA}
\affiliation{Center for Research and Exploration in Space Science and Technology II, NASA/GSFC, Greenbelt, MD 20771, USA}
\email{dana.r.louie@nasa.gov}

\author[0000-0001-5347-7062]{Joshua E. Schlieder}
\affiliation{Exoplanets and Stellar Astrophysics Laboratory (Code 667), NASA Goddard Space Flight Center, Greenbelt, MD 20771, USA}
\email{joshua.e.schlieder@nasa.gov}  

\author[0000-0002-4235-6369]{Robert F. Wilson}
\affiliation{Department of Astronomy, University of Maryland, College Park, MD 20742, USA}
\affiliation{Exoplanets and Stellar Astrophysics Laboratory (Code 667), NASA Goddard Space Flight Center, Greenbelt, MD 20771, USA}
\affiliation{Center for Research and Exploration in Space Science and Technology II, NASA/GSFC, Greenbelt, MD 20771, USA}
\email{robert.f.wilson@nasa.gov}  

\author[0000-0003-0430-3335]{Evan Bray}
\affiliation{KBR, Greenbelt, MD 20771, USA}
\email{Evan.Bray@nasa.gov}

\author{Scott Rohrbach}
\affiliation{Optics, Photonics, and Lasers Branch (Code 522), NASA Goddard Space Flight Center, Greenbelt, MD 20771, USA}
\email{scott.rohrbach@nasa.gov}  

\collaboration{all}{The Nancy Grace Roman Telescope}

\begin{abstract}
The Nancy Grace Roman Space Telescope launched on August 30th, 2026 and is expected to begin its science mission in 2027. Although Roman's core science will focus primarily on faint and distant sources, as a survey telescope, it will also observe many bright sources that can saturate the detector pixels. Our goal is to study the behavior caused by bright sources dispersed through the grism and to assess how detector properties may impact scientific analysis. We analyze detector saturation behavior in Roman's Wide Field Instrument (WFI) using Thermal Vacuum Test 2 (TVAC2) bright-source grism data. When bright sources are dispersed through the grism, we find that the faint and diffuse combined orders cause significant structured stray light. We also find that pixels neighboring saturated pixels continue to exhibit linear growth behavior before reaching saturation limits, in contrast to the non-linearity reported for non-dispersed bright point sources.  
\end{abstract}

\section{Introduction} 
The Nancy Grace Roman Space Telescope (Roman) is NASA's next flagship mission, designed to conduct wide-field survey science spanning cosmology, exoplanets, and other areas of astrophysics. The key enabler of Roman's survey mission is the Wide Field Instrument (WFI), which is equipped with 18 Teledyne H4RG-10 sensor chip assembles (SCAs) arranged in a 6 x 3 array \citep{Schlieder_2024, Mosby_2020, Mosby_2025}. WFI contains an Element Wheel Assembly (EWA)  with 8 filters, the P127 prism (0.75 -- 1.80 $\mu$m), the G150 grism (1.0 -- 1.93 $\mu$m), and a Dark element. The grism enables slitless spectroscopy by dispersing light from sources across the WFI field of view into spectral traces \citep{Bray_2024}. We subjected the grism to bright saturating sources during WFI’s second Thermal Vacuum Test (TVAC2), conducted in the Titan chamber at BAE Space \& Mission Systems in Boulder, CO, from March to May 2024. In this research note, we analyze how saturation effects in the H4RG-10 detectors manifest during grism observations and we make additional plots available on Zenodo (\href{https://doi.org/10.5281/zenodo.22214576}{DOI: 10.5281/zenodo.22214576})

\section{Methods}

\label{sec:style}
During TVAC2, we used a NASA Goddard Space Flight Center-developed telescope simulator called the Stimulus of Ray Cones \citep[SORC;][]{Schlieder_2024,Louie_2026} to project saturating point sources through the grism and onto SCAs 2 ($\sim$0.9 mag), 5 ($\sim$3.7 mag), 8 ($\sim$7.8 mag), and 14 ($\sim$-2.3 mag). We used the SORC SPLIT-IR input described in \cite{Louie_2026} to assess bright-source effects from the grism's diffuse combined-order structure \citep{Bray_2024} and investigate pixel-level saturation. Initial data processing and calibrations 
are detailed in \cite{Louie_2026}. We used the following methodology:

\begin{itemize}

    \item Applied the Improved Roman Reference Correction to mitigate correlated $1/f$ noise and removed detector-dependent structure from individual frames.

    \item Subtracted a superbias frame constructed from the corresponding bright-star dark exposures to remove the detector bias while preserving the source signal \citep{2024romanrept..673B}.

    \item Applied a per-pixel gain conversion to convert the detector signal from analog-to-digital units to electrons, accounting for variations in the response of individual pixels \citep{Rauscher_2007}.
\end{itemize}

We first visually examined the effects of saturation through the grism, then characterized saturation at the pixel level by examining the up-the-ramp evolution of individual pixels throughout each exposure. Because the per-pixel gain correction produces slightly different effective saturation levels, 
we identified saturated pixels using the raw detector frames and normalized the gain-corrected signal of each pixel by its maximum value. We then tracked neighboring pixels across the ramp to compare their signals as nearest neighbors approached and reached saturation, following the procedure described in \cite{Louie_2026}.

\section{Results} \label{sec:style}
\begin{figure}[h]      

    \centering   
        \makebox[\textwidth][c]{%
        \includegraphics[width=1.15\textwidth]{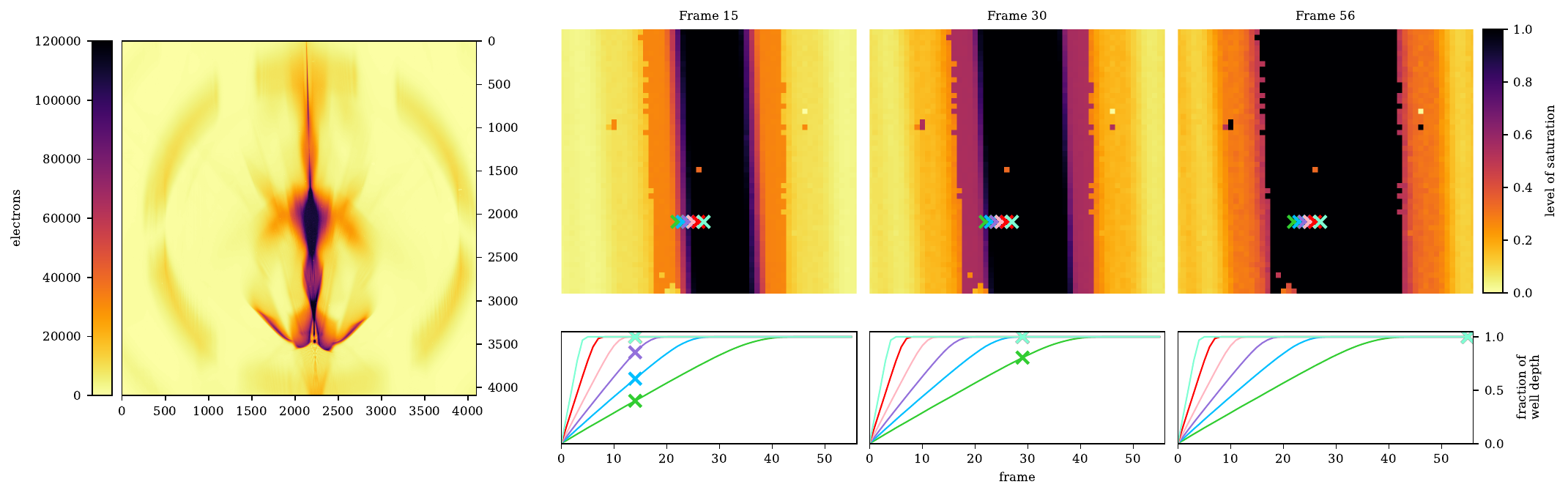}
    }

    \caption{(Left) View of the final frame of the SCA 5, 3.4 mag data. The Roman grism disperses light into multiple diffraction orders (e.g., (0,0), (1,1)), with most of the light concentrated in the primary (1,1) order used for science spectra. The extreme brightness of the simulated source reveals the diffuse structure of the additional diffraction orders surrounding the saturated (1,1) trace. The color scale is normalized to the detector saturation limit. The faint, broken ring surrounding the grism structure is scattered light from the telescope simulator and is not expected in flight data. (Right) Progression of saturation in the SCA 8, 7.8 mag data. The top panels show zoomed-in images of the primary (1,1) grism trace at frames 15, 30, and 56, from left to right. Colored ``X'' symbols mark neighboring pixels spaced one pixel apart, and the corresponding bottom panels show the accumulation of charge in these pixels throughout the exposure. The ``X'' symbols on the curves indicate the charge accumulated at the frame shown above. Pixels close to the center of the trace saturate very quickly, in only a few frames and then the extent of saturation grows over time as more charge is accumulated. A pixel reaches saturation at the point where its curve plateaus.}

    \label{fig:fig1}       
\end{figure}

 Saturation first appears along the central (1,1) grism trace and subsequently expands outward pixel-by-pixel along the dispersed spectrum 
 ultimately revealing the full spatial distribution of flux imparted by the grism. 
 The combined grism orders produce a faint, diffuse structure surrounding the central dispersed spectrum (Figure~\ref{fig:fig1},\textit{~left}), with its spatial extent increasing with source brightness. 
The brighter sources saturated substantially more pixels: by frame 56, the $-2.3$ mag source had saturated $\sim$16 million pixels, compared to $\sim$5 million, $\sim$300 thousand, and $\sim$13 thousand pixels for the 0.9, 3.4, and 7.8 mag sources, respectively (Figure~\ref{fig:fig1},\textit{~right}). 
 
 For bright sources dispersed through the grism, we observe that up-the-ramp signal integration curves for pixels neighboring saturated pixels show similar linear growth prior to their own saturation (Figure~\ref{fig:fig1},\textit{~right}). The linear growth of saturation we observed for bright sources dispersed through the grism differed from the non-dispersed bright point-source saturation analyzed by \citet{Louie_2026}.
 This suggests minimal cross-pixel coupling effects in the tested flux regimes, which may be driven by the dispersion of bright light over many pixels, rather than being focused into the core of the point spread function. 

\section{Conclusion} \label{sec:style}
When bright sources are dispersed through the WFI grism, they result in a bright, saturated central trace, surrounded by diffuse structure due to combined orders inherent to the optical design (Figure~\ref{fig:fig1},\textit{~left}). We also find that brighter simulated sources reach saturation more rapidly than fainter sources, with charge accumulation proceeding approximately linearly in individual pixels until approaching the full-well limit, where the signal plateaus. We do not observe strong evidence for the significant cross-pixel coupling prior to saturation seen for non-dispersed point sources in~\cite{Louie_2026}. When dispersed through the grism, pixels neighboring saturated pixels exhibit linear growth behavior before reaching their respective saturation limits, without additional non-linearity.  These results indicate that the dispersion of light plays an important role in how bright-source systematics manifest in WFI observations. This early characterization of saturation in Roman WFI grism data informs strategies for mitigating bright-source systematics and using bright and saturated sources for science in upcoming Roman surveys.

\begin{acknowledgments}
LC thanks DL, JS, RW for their mentorship. LC also thanks NASA OSTEM and the Roman Collaboration for this opportunity. The authors acknowledge the use of the Advanced Data Analytics Platform Technologies (ADAPT) Science Cloud and its GPU PRISM cluster provided by the NASA Center for Climate Simulation (NCCS) at the Goddard Space Flight Center. The work is supported by NASA under award number 80GSFC24M0006. 
\end{acknowledgments}

\newpage

\bibliography{sample701.bib}

@article{Mosby_2020,
   title={Properties and characteristics of the Nancy Grace Roman Space Telescope H4RG-10 detectors},
   volume={6},
   ISSN={2329-4124},
   url={http://dx.doi.org/10.1117/1.JATIS.6.4.046001},
   DOI={10.1117/1.jatis.6.4.046001},
   number={04},
   journal={Journal of Astronomical Telescopes, Instruments, and Systems},
   publisher={SPIE-Intl Soc Optical Eng},
   author={Mosby, Gregory and Rauscher, Bernard J. and Bennett, Chris and Cheng, Edward S. and Cheung, Stephanie and Cillis, Analia and Content, David and Cottingham, Dave and Foltz, Roger and Gygax, John and Hill, Robert J. and Kruk, Jeffrey W. and Mah, Jon and Meier, Lane and Merchant, Chris and Miko, Laddawan and Piquette, Eric C. and Waczynski, Augustyn and Wen, Yiting},
   year={2020},
   month=nov }

@TECHREPORT{2024romanrept..673B,
       author = {{Betti}, S.~K. and {Casertano}, S. and {Brandt}, T. and {Cosentino}, R. and {Desjardins}, T. and {Sanchez}, J.},
        title = "{Application and Characterization of IRRC on Roman/WFI TVAC1 Data}",
  institution = {STScI},
         year = 2024,
       number = {Technical Report Roman-STScI-000673},
 howpublished = {Nancy Grace Roman Space Telescope Technical Reports, 16 pp.},
       adsurl = {https://ui.adsabs.harvard.edu/abs/2024romanrept..673B}
}

@article{Rauscher_2007,
   title={Detectors for theJames Webb Space TelescopeNear‐Infrared Spectrograph. I. Readout Mode, Noise Model, and Calibration Considerations},
   volume={119},
   ISSN={1538-3873},
   url={http://dx.doi.org/10.1086/520887},
   DOI={10.1086/520887},
   number={857},
   journal={Publications of the Astronomical Society of the Pacific},
   publisher={IOP Publishing},
   author={Rauscher, Bernard J. and Fox, Ori and Ferruit, Pierre and Hill, Robert J. and Waczynski, Augustyn and Wen, Yiting and Xia‐Serafino, Wei and Mott, Brent and Alexander, David and Brambora, Clifford K. and Derro, Rebecca and Engler, Chuck and Garrison, Matthew B. and Johnson, Thomas and Manthripragada, Sridhar S. and Marsh, James M. and Marshall, Cheryl and Martineau, Robert J. and Shakoorzadeh, Kamdin B. and Wilson, Donna and Roher, Wayne D. and Smith, Miles and Cabelli, Craig and Garnett, James and Loose, Markus and Wong‐Anglin, Selmer and Zandian, Majid and Cheng, Edward and Ellis, Timothy and Howe, Bryan and Jurado, Miriam and Lee, Ginn and Nieznanski, John and Wallis, Peter and York, James and Regan, Michael W. and Hall, Donald N. B. and Hodapp, Klaus W. and Böker, Torsten and De Marchi, Guido and Jakobsen, Peter and Strada, Paolo},
   year={2007},
   month=jul, pages={768–786} }

@ARTICLE{Louie_2026,
       author = {{Louie}, Dana R. and {Wilson}, Robert F. and {Barclay}, Thomas and {Schlieder}, Joshua E. and {Bond}, Nicholas and {Bray}, Evan and {Cabrera}, Mario and {Cheung}, Stephanie and {Choi}, Ami and {Cillis}, Analia and {Collins}, Nicholas and {Groff}, Tyler D. and {Hill}, Robert J. and {Kruk}, Jeffrey and {Mosby}, Gregory and {Paine}, Jennie and {Rauscher}, Bernard J. and {Reichard}, Timothy A. and {Rizzo}, Maxime and {Rohrbach}, Scott and {Schanche}, Nicole and {Switzer}, Eric R.},
        title = "{Nancy Grace Roman Space Telescope Wide Field Instrument: Bright Point Source Saturation Response and Persistence Properties from Thermal-Vacuum Testing}",
      journal = {arXiv e-prints},
         year = 2026,
        month = jul,
          eid = {arXiv:2607.18419},
        pages = {arXiv:2607.18419},
          doi = {10.48550/arXiv.2607.18419},
archivePrefix = {arXiv},
       eprint = {2607.18419},
 primaryClass = {astro-ph.IM},
       adsurl = {https://ui.adsabs.harvard.edu/abs/2026arXiv260718419L}
}

@ARTICLE{Mosby_2025,
       author = {{Mosby}, Gregory and {Hill}, Bob and {Cillis}, Analia and {Foltz}, Roger and {Bennett}, Chris and {Cabrera}, Mario and {Hickey}, Michael and {Merchant}, Chris and {Waczynski}, Augustyn and {Wen}, Yiting and {Rauscher}, Bernie and {Wollack}, Edward J. and {Kruk}, Jeff and {Auyeung}, John and {Schlieder}, Joshua and {Switzer}, Eric and {Rizzo}, Maxime and {Zimmerman}, Neil and {Cheng}, Edward and {Cabelli}, Craig and {Piquette}, Eric and {Pan}, Jianmei and {Farris}, Mark and {Petersen}, Anders and {Gygax}, John and {Stuart}, Colin and {Cottingham}, Dave and {Mah}, Jonathan and {Cheung}, Stephanie and {Meier}, Lane and {Kelly}, Daniel and {Mann}, Steven},
        title = "{Summary of the Nancy Grace Roman Space Telescope flight detector performance}",
      journal = {Journal of Astronomical Telescopes, Instruments, and Systems},
         year = 2025,
        month = jan,
       volume = {11},
          eid = {011210},
        pages = {011210},
          doi = {10.1117/1.JATIS.11.1.011210},
       adsurl = {https://ui.adsabs.harvard.edu/abs/2025JATIS..11a1210M}
}

@INPROCEEDINGS{Schlieder_2024,
       author = {{Schlieder}, Joshua E. and {Barclay}, Thomas and {Barnes}, Amethyst and {Bray}, Evan and {Choi}, Ami and {Cromey}, Benjamin and {Delker}, Thomas and {Finch}, Timothy and {Frater}, Eric H. and {Hill}, Robert J. and {Kruk}, Jeffrey and {Lasco}, Jeffrey and {Louie}, Dana R. and {Malhotra}, Sangeeta and {McEnery}, Julie E. and {Mosby}, Gregory and {Paine}, Jennie and {Perkins}, Jeremy S. and {Rauscher}, Bernard J. and {Rhoads}, James E. and {Rizzo}, Maxime and {Sabatke}, Derek and {Schweickart}, Rusty and {Shukis}, Diana and {Switzer}, Eric R. and {Wollack}, Edward J. and {Zellem}, Robert T. and {Zimmerman}, Neil T.},
        title = "{Survey science with the Nancy Grace Roman Space Telescope Wide Field Instrument}",
    booktitle = {Space Telescopes and Instrumentation 2024: Optical, Infrared, and Millimeter Wave},
         year = 2024,
       editor = {{Coyle}, Laura E. and {Matsuura}, Shuji and {Perrin}, Marshall D.},
       series = {Society of Photo-Optical Instrumentation Engineers (SPIE) Conference Series},
       volume = {13092},
        month = aug,
          eid = {130920S},
        pages = {130920S},
          doi = {10.1117/12.3020622},
       adsurl = {https://ui.adsabs.harvard.edu/abs/2024SPIE13092E..0SS}
}

@ARTICLE{Bray_2024,
       author = {{Bray}, Evan and {Batkis}, Mateo and {Chambers}, Victor J. and {Dominguez}, Margaret and {Eegholm}, Bente and {Gao}, Guangjun and {Gong}, Qian and {Halliday}, Wesley and {Howe}, Elias and {Kruk}, Jeffrey and {Malumuth}, Eliot and {Malhotra}, Sangeeta and {Marx}, Catherine and {Rhoads}, James and {Rizzo}, Maxime and {Schlieder}, Joshua E. and {Seide}, Laurie and {Switzer}, Eric R. and {Voris}, Jay},
        title = "{Spectral characterization of the Grism and Prism slitless spectrometers for the Nancy Grace Roman Space Telescope}",
      journal = {Journal of Astronomical Telescopes, Instruments, and Systems},
         year = 2024,
        month = jan,
       volume = {10},
          eid = {014003},
        pages = {014003},
          doi = {10.1117/1.JATIS.10.1.014003},
       adsurl = {https://ui.adsabs.harvard.edu/abs/2024JATIS..10a4003B}
}
\bibliographystyle{aasjournalv7}

\end{document}